# Strongly enhanced current densities in $Sr_{0.6}K_{0.4}Fe_2As_2$+Sn superconducting tapes

He Lin,[1] Chao Yao,[1] Xianping Zhang,[1] Haitao Zhang,[1] Dongliang Wang,[1] Qianjun Zhang,[1] Yanwei Ma,[1,*] Satoshi Awaji,[2] and Kazuo Watanabe[2]

[1] *Key Laboratory of Applied Superconductivity, Institute of Electrical Engineering, Chinese Academy of Sciences, PO Box 2703, Beijing 100190, China*
[2] *High Field Laboratory for Superconducting Materials, Institute for Materials Research, Tohoku University, Sendai 980-8577, Japan*

(*) Author to whom any correspondence should be addressed. E-mail: ywma@mail.iee.ac.cn

**Abstract**

Improving transport current has been the primary topic for practical application of superconducting wires and tapes. However, the porous nature of powder-in-tube (PIT) processed iron-based tapes is one of the important reasons for low critical current density ($J_c$) values. In this work, the superconducting core density of *ex-situ* $Sr_{0.6}K_{0.4}Fe_2As_2$+Sn tapes, prepared from optimized precursors, was significantly improved by employing a simple hot pressing as an alternative route for final sintering. The resulting samples exhibited optimal critical temperature ($T_c$), sharp resistive transition, small resistivity and high Vickers hardness ($Hv$) value. Consequently, the transport $J_c$ reached excellent values of $5.1 \times 10^4$ A/cm$^2$ in 10 T and $4.3 \times 10^4$ A/cm$^2$ in 14 T at 4.2 K, respectively. Our tapes also exhibited high upper critical field $H_{c2}$ and almost field-independent $J_c$. These results clearly demonstrate that PIT pnictide wire conductors are very promising for high-field magnet applications.

Since the discovery of iron-based superconductors[1], the development of wire and tape preparation for practical application was rapidly followed[2-4]. Particularly, 122-type iron-pnictides are most potentially useful due to their relatively high $T_c$ up to 38 K[5], ultrahigh $H_{c2}$ above 100 T[6, 7] and small anisotropy γ about 1.5-2[8]. A commonly fabricating route for 122-type wires and tapes is *ex-situ* powder-in-tube (PIT) process because of its simplicity and low-cost production. However, PIT pnictide wires and tapes exhibit intrinsic weak-link behavior of grain boundaries and extrinsic factors which induce poor connection between polycrystalline grains and thus low global $J_c$ values. Various methods including chemical addition[9-13], rolling texture[3, 13], hot isostatic pressing[14], cold pressing[15] and cycles of cold deformation and heat treatment[16] have been dedicated to improve the superconducting properties of pnictide wires and tapes. For instance, by using a texturing process plus Sn addition to improve grain connectivity, high transport $J_c$ values were achieved in Fe-clad $Sr_{1-x}K_xFe_2As_2$ (Sr-122) tapes[13]. High-purity and nearly 100% dense $Ba_{1-x}K_xFe_2As_2$ (Ba-122) cores could be achieved by applying HIP technique and low-temperature sintering[14]. Recently, Togano *et al.* reported a large $J_c$ enhancement by cold pressing technology, which was attributed to densification and the change of crack structure[16]. In past five years, the critical current densities of 122-type wires and tapes sharply increased from zero to above $10^4$ A/cm$^2$ (4.2 K, 10 T), indicating its promising future for high-field applications. However, these transport $J_c$ reported so far are still lower than the practical level, because polycrystal superconductors suffer from the disadvantages of impurity phase, inhomogeneity, low density and crack. Therefore, further $J_c$ enhancement of PIT tapes may be realized by optimizing precursor powders and increasing core density.

High-quality precursor powders are vital to PIT pnictide process[14, 17-19]. Any changes in the composition or homogeneity of starting precursors could affect the superconducting performances of pnictide tapes. Our recent work already suggested that two-step sintering process of precursors helped to complete reaction and homogenous microstructure of Ag-added Sr-122 tapes whose $J_c$ was greatly increased by an order of magnitude[20]. On the other hand, hot pressing process could increase core density and improve grain alignment in Bi-2223 tapes, which in return led to higher zero-field $J_c(H=0)$ and in-field $J_c(H)$[21, 22]. The $J_c$ values of $MgB_2$ tapes were also strongly improved by hot pressing, due to much denser superconducting cores[23, 24]. The external pressure at high temperature can reduce the tape volume and suppress the introduction of voids during the reaction process of the $MgB_2$ synthesis. Therefore, we expected that the porous nature of pnictide tapes, such as low density, cracks and voids, would be effectively solved by the hot pressing process.

Here we fabricated Ag-clad $Sr_{1-x}K_xFe_2As_2$+Sn tapes through two-step sintering process of optimized precursors (2S precursors) and hot pressing process (HP process). A high transport current to $J_c$ (4.2 K, 10 T) ≈ $5.1 \times 10^4$ A/cm$^2$ is observed. Such a phenomenon can be qualitatively explained by the combination of improved phase formation and high core density. In addition, the conventional precursors (1S precursors) and ordinary rolled tapes (OR tapes) were also fabricated for comparison.

**Results**

Fig. 1 displays the magnetization versus temperature (M-T) curves of conventional 1S precursor powders and optimized 2S precursor powders. The striking result is that the diamagnetic response of 2S precursors at low temperature (below 15 K) is about two times higher than that of 1S precursors. This indicates that the second sintering process of 2S precursors with Sn addition facilitates the formation of Sr-122 superconducting phase. Nevertheless, onset $T_c$ decreases slightly from 36.0 K for 1S precursors to 34.7 K for 2S precursors.

The X-ray diffraction (XRD) patterns of various Sr-122 tapes prepared from 1S and 2S precursors are shown in Fig. 2. All four-type samples exhibit a well ThCr$_2$Si$_2$-type structure despite of the Ag peaks which are contributed from the Ag sheath, ensuring that a single-phase Sr$_{1-X}$K$_X$Fe$_2$As$_2$ superconductor is obtained in final tapes prepared from both precursors. M-T curves of these Sr-122 tapes were measured and the results are reported in Fig. 3, where all samples show a sharp transition. Compared with OR-1S and HP-1S samples, both OR-2S and HP-2S tapes produced from optimized 2S precursors exhibit an enhanced diamagnetic response at the temperature below $T_c$, which is consistent with the M-T observation of precursors. The onset $T_c$ values for OR-1S, HP-1S, OR-2S and HP-2S tapes are 31.8, 32.4, 34.1 and 33.8 K, respectively. It is interesting that onset $T_c$ is very susceptible to different precursors but hardly affected by hot pressing. Regardless of the external pressure, Onset $T_c$ of OR-1S and HP-1S tapes are about 4 K lower than that of 1S precursors. We speculated that some non-uniform phases still remain in 1S precursors by one-step sintering process[4, 14, 17]. In contrast, the onset $T_c$ values of OR-2S and HP-2S tapes display slight decrease (~0.8 K) after tape fabrication with respect to 2S precursors. The magnetic results suggest that two-step sintering process of precursors improves the homogeneity and grain crystallization of final Sr-122 superconducting phases.

In addition, the inset of Fig. 3 shows resistivity versus temperature (R-T) curves for OR-2S and HP-2S tapes. Similar onset $T_c$ values ~ 35.3 K are observed for both tapes. Relatively, the resistivity of OR-2S and HP-2S tapes drops to zero $T_c$ at 34.5 and 34.7 K, respectively. The sharp resistivity transitions (< 1K) indicate well electromagnetic homogeneity of our samples[19, 25, 26]. EDS mapping as shown in Fig. 4 further demonstrates that the elements Sr, K, Fe and As of superconducting phase homogeneously distributed throughout the superconducting core in 2S tapes (the measured stoichiometry is $Sr_{0.61}K_{0.41}Fe_2As_{2.02}$). For comparison, 1S tapes show superconducting elements enriched in some areas (supplemental Fig. S1). Obviously, the resistivity of HP-2S tape is much smaller than the value of OR-2S sample, giving strong evidence of the improvement of the grain connectivity. Magneto-resistivity curves in applied fields from 0 to 9 T of the above two samples are also measured (supplemental Fig. S2). With increasing magnetic field, the superconducting transitions of both samples shift towards lower temperature. The upper critical field $H_{c2}$ is evaluated with the criteria of 90% of normal state resistivity respectively. The $H_{c2}$ at zero-temperature can be deduced using the Werthamer–Helfand–Hohenberg (WHH) formula: $H_{c2}(0)= -0.693T_c(dH_{c2}/dT)$. For the OR-2S and HP-2S samples, taking $T_c$ = 35.3 K, the values of $H_{c2}(0)$ are as high as 115 and 110 T.

We now turn our attention to the effects of hot pressing on final tapes. Fig. 5

shows the cross-section images of Sr-122 tapes with and without hot pressing (HP-2S and OR-2S). It is clear that hot pressing significantly reduces the tape thickness from 0.44 mm to 0.3 mm. The cross-sectional area of the superconducting core in hot-pressed tapes is 0.28 mm$^2$, decreased by around 30% compared to that of the rolled tapes. Similar result was also reported in the previous PIT processed MgB$_2$ tapes[23]. Vickers hardness ($Hv$) measurement was carried out for both tape samples to further investigate the change of core density. The average $Hv$ values of the OR-2S and HP-2S tapes are 61.2 and 97.3, respectively, proving that hot pressing leads to a much denser core. However, the $Hv$ values are still lower for Sr-122 tapes than previously reported Ba-122 tapes[16], which may be caused by the different processing conditions.

Fig. 6 shows the $J_c$-B properties at 4.2 K of Sr-122 tapes with and without hot pressing. Only data above 2 T are given, because $I_c$ at lower field region was too high to be measured. For the tapes without hot pressing, the $J_c$ value of 2.0 ×10$^4$ A/cm$^2$ in 10 T at 4.2 K are obtained for OR-2S tapes, which is about 34% higher than that of OR-1S tapes. The striking result of Fig. 6 is that both hot-pressed tapes (HP-1S and HP-2S) show a great enhancement of $J_c$ values in the whole field up to 14 T. The transport $J_c$ of HP-2S tapes reaches 5.1 ×10$^4$ A/cm$^2$ in 10 T at 4.2 K. Our Sr$_{0.6}$K$_{0.4}$Fe$_2$As$_2$+Sn tapes also exhibit almost field-independent $J_c$ ($H$) compared to Nb$_3$Sn, NbTi and MgB$_2$ superconductors. The $J_c$ of HP-2S samples still maintains a high value of 4.3 ×10$^4$ A/cm$^2$ in 14 T, which is obviously going to be true as Sr-122 has a high upper critical field $H_{c2}$(0) (≥110 T). Because of such high $J_c$-H performance, it is convincible that iron-pnictide has a very promising future for practical high-field applications.

Further support to the existing $J_c$ improvement has been obtained from the analysis of the microstructural difference between rolled and hot-pressed tapes. Figs. 7 (a-f) show typical SEM images of the superconducting core of OR-2S and HP-2S tapes. The in-plane images (a-b) and longitudinal-section images (e-f) were performed on the core surface after peeling off the Ag sheath. The samples in images (c-d) were polished carefully so that the level of porosity could be quantified. Both tape samples have well-developed grains. However, the OR-2S samples have loose microstructure with many voids and residual cracks generated during packing and cold deformation (Figs. 7 (a) and (c)), and thus the grains are not well connected. Some small crushed particles are also observed and the grain sizes vary around 2-5 μm. In contrast, as presented in Figs. 7 (b) and (d), the pores and cracks almost disappear in whole core area of HP-2S tapes, resulting in a much higher density and accordingly better connectivity between the Sr-122 grains. It should be noted that cold pressing easily introduced the crack structures which run transverse to the tape length[16]. However, similarly large cracks are absent in our hot-pressed samples, suggesting that hot deformation can eliminate defects effectively. The average size of plate-like grains is about 4-5 μm, which is slightly larger than that of OR-2S samples. This more uniform microstructure could further improve the grain coupling. Figs. 7(e) and (f) show the longitudinal microstructure of OR-2S and HP-2S samples, respectively. Due to the two dimensional nature of crystal structure of 122 phase[7], most Sr-122 grains present a planar structure in rolled tapes (Fig. 7(e)). However, some degree of grain twisting is obviously observed in hot-pressed tapes (Fig. 7(f)), which can be due to the softness of grains during hot deformation. The bending of grains could prevent the formation of cracks effectively in HP-2S samples, introducing greatly enhanced strong-links[21, 22, 27]. Clearly, the hot pressing process in this work significantly increases the core density of pnictide tapes without crushing of grains, and thus resulting in enhanced transport properties.

On the other hand, Sr-122 grains are oriented along with the tape axis marked by white arrows in Figs. 7 (e) and (f), meaning that the grain coupling in our samples is strengthened by the simple deformation technique. This result is also confirmed by XRD analysis as shown in Fig. 2. The degree of grain alignment can be quantified by the Lotgering method as follows[28]. The texture parameter $F= (\rho-\rho_0)/(1-\rho_0)$, where $\rho=\Sigma I(00l)/I(hkl)$, $\rho_0=\Sigma I_0(00l)/I_0(hkl)$. $I$ and $I_0$ are the intensities of each reflection peak (hkl) for the oriented and random samples, respectively. The F values for OR-1S, OR-2S, HP-1S and HP-2S samples are 0.32, 0.35, 0.35 and 0.31, respectively. The degree of grain alignment for all samples is lower than that of Fe-clad Sr-122 tapes[13, 15] because of using low hardness of Ag sheath. However, the F values almost remain constant after hot pressing, suggesting that the hot pressing process in this work has less effect on the degree of *c*-axis texture, because the relatively low hot pressure (~10 MPa) seems not enough to affect the grain alignments. Possibly, a further enhancement of texture may be expected with the optimization of the hot pressing conditions.

**Discussion**

High performance Ag-clad Sr-122 tapes have been successfully fabricated by *ex-situ* PIT method, involving the two-step intermediate sintering process of precursors and hot pressing technique. It is thought that the improved Sr-122 phase formation and high core density are responsible for the significant $J_c$ enhancement.

Firstly, controlling the properties of precursor powders is vital to the final performance of superconductors. The precursors obtained by first conventional sintering step have large grains with few impurities[17]. The second annealing treatment for 2S precursors allows more Sr-122 superconducting phase to form through solid-state diffusion[25]. Then final OR-2S and HP-2S tapes show optimal $T_c$, enhanced diamagnetism and sharp resistivity transition, ensuring that the improved Sr-122 phase formation has been obtained in both 2S tapes with respect to 2S precursors. Therefore, both 2S tapes exhibit better transport properties in comparison with the standard 1S tapes.

Secondly, early studies showed that the strong granularity of polycrystalline compounds could cause a strong limitation to the global current flow[29, 30]. Indeed, the porous nature of PIT pnictide tapes has extremely negative effect on transport $J_c$ values. By employing simple hot pressing technique as an alternative route for final sintering, we are able to obtain quite high dense superconducting core, which is a main reason for superior $J_c$ performance as shown in Fig. 6. It is found that hot pressing is very effective to reduce the pores and cracks formed by the previous packing and cold deformation. This densification contributes to well grain connectivity in the whole core area. On the other hand, although the superconducting cross sections of HP tapes are greatly reduced compared with those of rolled tapes, the bending of Sr-122 grains without large crack structure is observed in the hot-pressed cores, thus becoming profitable for strong-links of grains[21, 22, 27]. This is in contrast to the result of cold-pressed tapes in which some crushed planar grains and large crack structures are easily introduced by strong cold force[15, 16]. The HP tapes exhibit small resistivity, high $Hv$ value and uniform microstructure. Accordingly, the remarkable $J_c$ values of $5.1 \times 10^4$ A/cm$^2$ in 10 T and $4.3 \times 10^4$ A/cm$^2$ in 14 T at 4.2 K have been achieved in our $Sr_{0.6}K_{0.4}Fe_2As_2$+Sn tapes. Based on the recent results, it is believed that further improvement in $J_c$ can be expected upon optimizing process parameters.

*Note added:* More recently, the wire and tape preparation for iron-pnictides was rapidly developed. NIMS group reported high transport $J_c$ in Ba-122 thin tapes by applying cold pressing process[31]. Immediately after that, our group announced that the transport $J_c$ exceeded $10^5$ A/cm$^2$ at 4.2 K and 10 T for Sr-122 tapes, which achieved the practical desired level of critical current densities[32].

**Methods**

**Sample preparation.** Ag-clad $Sr_{0.6}K_{0.4}Fe_2As_2$ tapes with Sn addition were fabricated by the *ex-situ* PIT method. Sr fillings, K pieces, and Fe and As powders were used as staring materials with the nominal composition of $Sr_{0.6}K_{0.4}Fe_2As_2$. In order to compensate for the loss of elements during the sintering, the starting materials contains 25 wt% excess K and 5 wt% excess As[33]. The starting materials were mixed and ground for about 12 hours by ball-milling method, which was performed in a QM-1SP4 planetary ball mill (Nanjing NanDa Instrument Plant, China) under argon atmosphere (dry milling). The milling jars and the milling balls were made from stainless steel.

The as-milled powders were further processed to precursors and tapes. The optimized precursors (2S precursors) with Sn addition were prepared through two-step sintering process. Firstly, milled powders were heat treated at 900 ℃ for 35 h. The sintered bulk was ground into powders and mixed with 5 wt% Sn by hand. Secondly, these powders with Sn addition were sintered at 900 ℃ for 30 min again. Then the synthesized precursors were packed into Ag tubes with OD 8 mm and ID 5 mm. These tubes were sealed and then cold worked into tapes (about 0.44 mm thickness) by drawing and flat rolling. The ordinary rolled tapes (OR-2S tapes) were subsequently sintered at 850 ℃ for 30 min. For the hot-pressed tapes (HP-2S tapes), as-rolled tapes were sandwiched between two pieces of metal sheets and pressed at ~10 MPa at 850 ℃ for 30 min. Furthermore, the conventional precursors (1S precursors) were prepared by one-step sintering process ($900^oC/35h$)[12, 13]. The prepared precursor was added with 5 wt% Sn to improve the grain connectivity. For comparison, as-rolled and hot-pressed Sr-122 tapes (OR-1S and HP-1S tapes) were also fabricated by similar process. The detailed sample names of precursors and tapes are listed in Table I.

**Measurements.** Magnetization versus temperature curves were performed with a superconducting quantum interference device magnetometer (SQUID). Resistivity measurements were carried out using a PPMS system. The cross sections were mechanical polished by emery paper and lapping paper, and then the areas of superconducting cores were calculated by optical microscope images. Vickers hardness of the tape samples was measured on the polished cross sections with 0.05 kg load and 10 s duration in a row at the center of the superconducting cross section. The phase composition and microstructure was characterized by X-ray diffraction (XRD), energy dispersive X-ray spectroscopy (EDS) and scanning electron microscope (SEM). The polished samples for SEM analysis were carried out after the mechanical polishing by emery paper and then Ar ion polishing by reactive ion etching (RIE) system. The transport current $I_c$ at 4.2 K and its magnetic field dependence were measured by the standard four-probe method with a criterion of 1 μV/cm at the High Field Laboratory for Superconducting Materials (HFLSM) at Sendai. Then the critical current was divided by the cross section area of the superconducting core to get the critical current density $J_c$. For each set of tapes, $I_c$ measurement was performed on several samples to ensure the reproducibility.

**Acknowledgments**

This work is partially supported by the National '973' Program (grant No. 2011CBA00105) and the National Natural Science Foundation of China (grant Nos. 51172230, 51320105015 and 51202243).

**Author contributions**

Y. W. M. planed and directed the research. H. L. fabricated the tape samples and carried out resistivity, magnetization, XRD, EDS and SEM experiments. X. P. Z. and C. Y. did the high-field $I_c$ measurement. S. A. and K. W. helped with transport measurement. D. L. W and Q. J. Z. helped with the tape preparation. X. P. Z., C.Y. and H. T. Z. contributed to manuscript preparation. H. L. and Y. W. M. wrote the manuscript. All the authors contributed to discussion on the results for the manuscript.

**Additional information**

Competing financial interests: The authors declare no competing financial interests.

Table I The sample names of precursors and Sr-122 tapes.

| Precursors | Ordinary tapes (unpressed) | Hot pressed tapes |
|---|---|---|
| One-step sintered precursors (1S) | OR-1S | HP-1S |
| Two-step sintered precursors (2S) | OR-2S | HP-2S |

# Captions

Figure 1 Magnetization versus temperature curves of the conventional 1S precursor powders and the optimized 2S precursor powders.

Figure 2 XRD patterns for various Sr-122 tapes prepared from 1S and 2S precursors. The data were obtained after peeling off the Ag sheath. The peaks of $Sr_{0.6}K_{0.4}Fe_2As_2$ phase are indexed. The peaks of Ag are contributed from the Ag sheath.

Figure 3 Volumetric magnetic susceptibility as a function of temperature for various Sr122+Sn tapes with and without hot pressing. The magnetic response was evaluated by warming above $T_c$ after zero-field-cooling to 10 K and applying a field of 20 Oe parallel to the sample's length. Inset showing the resistivity versus temperature curves of rolled OR-2S sample and hot pressed HP-2S tape. All data were obtained after peeling off the Ag sheath.

Figure 4 EDS mapping images of the superconducting core of HP-2S sample.

Figure 5 Optical microscope images of the cross sections of (a) rolled OR-2S tape and (b) hot-pressed HP-2S tape.

Figure 6 Magnetic field dependence of transport $J_c$ at 4.2 K for various Sr-122 tapes with and without hot pressing. The applied fields up to 14 T were parallel to the tape surface.

Figure 7 SEM images showing planar views of (a) rolled OR-2S sample, (b) hot-pressed HP-2S sample, (c) polished OR-2S sample and (d) polished HP-2S sample; longitudinal sections of (e) rolled OR-2S sample and (f) hot-pressed HP-2S tape. The tape axis has been labeled using white arrows.

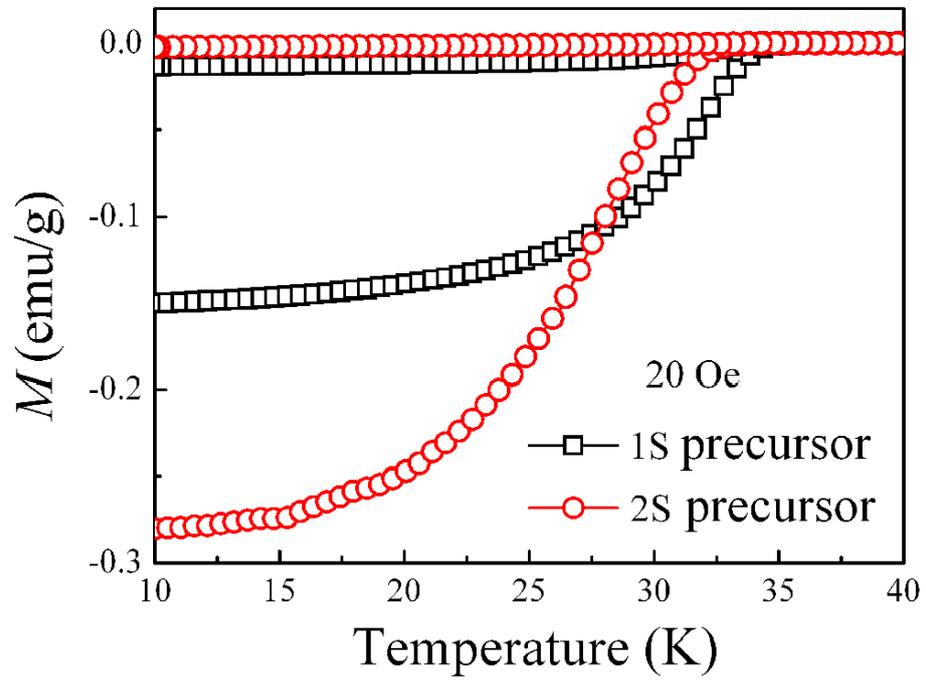

Figure 1 Lin et al.

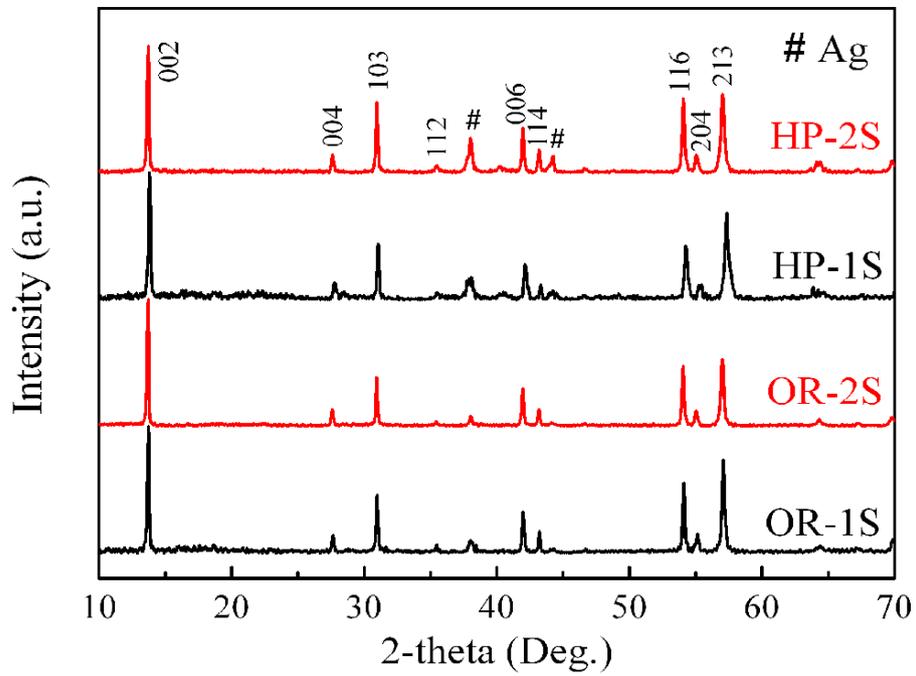

Figure 2 Lin et al.

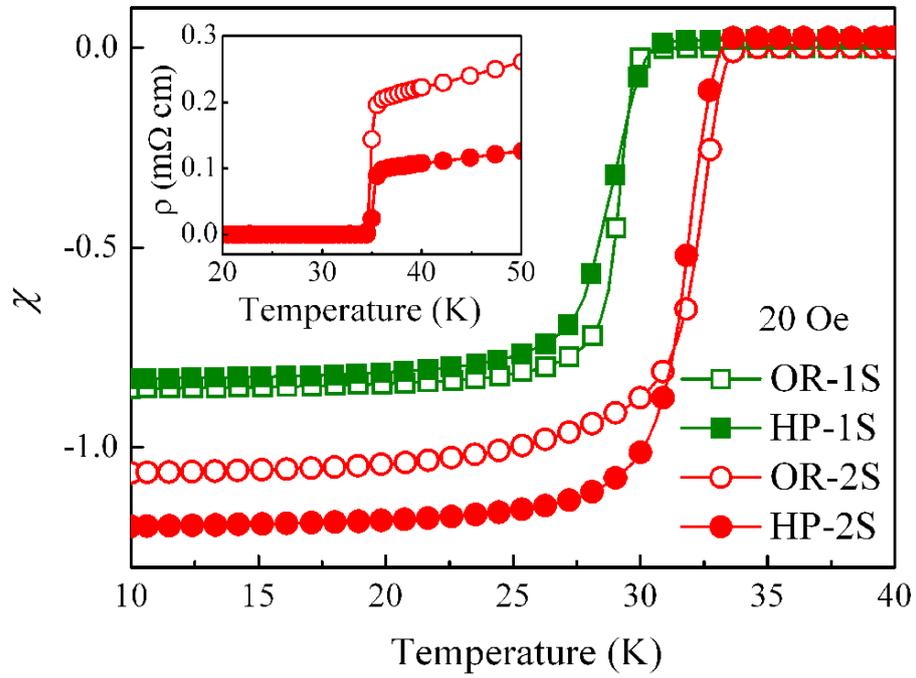

Figure 3 Lin et al.

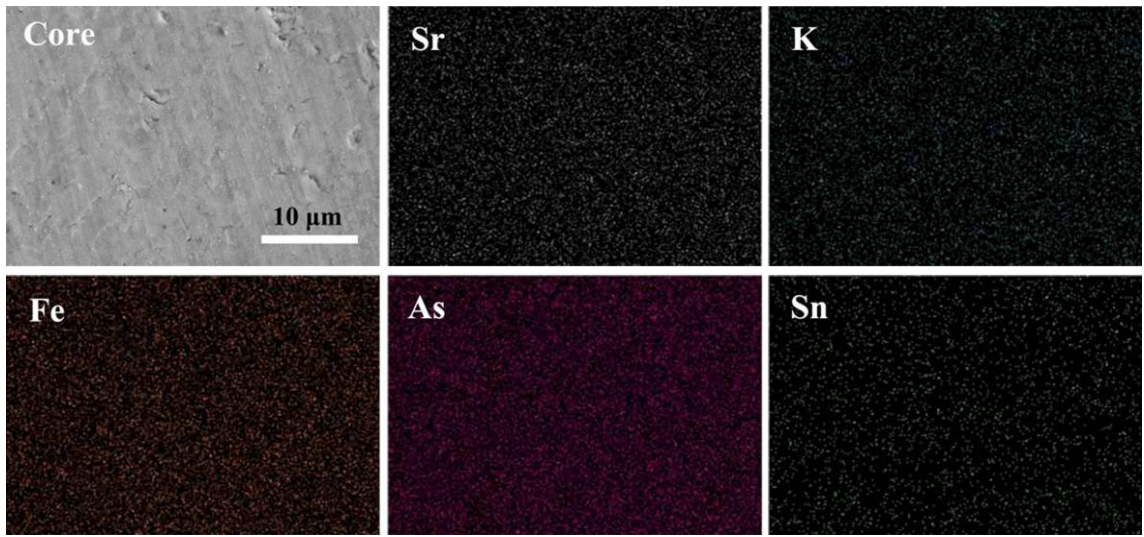

Figure 4 Lin et al.

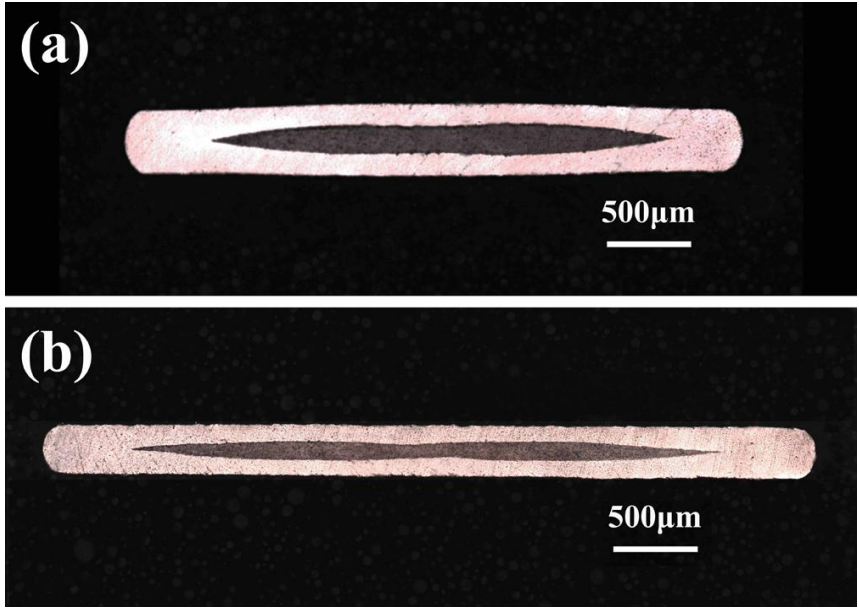

Figure 5 Lin et al.

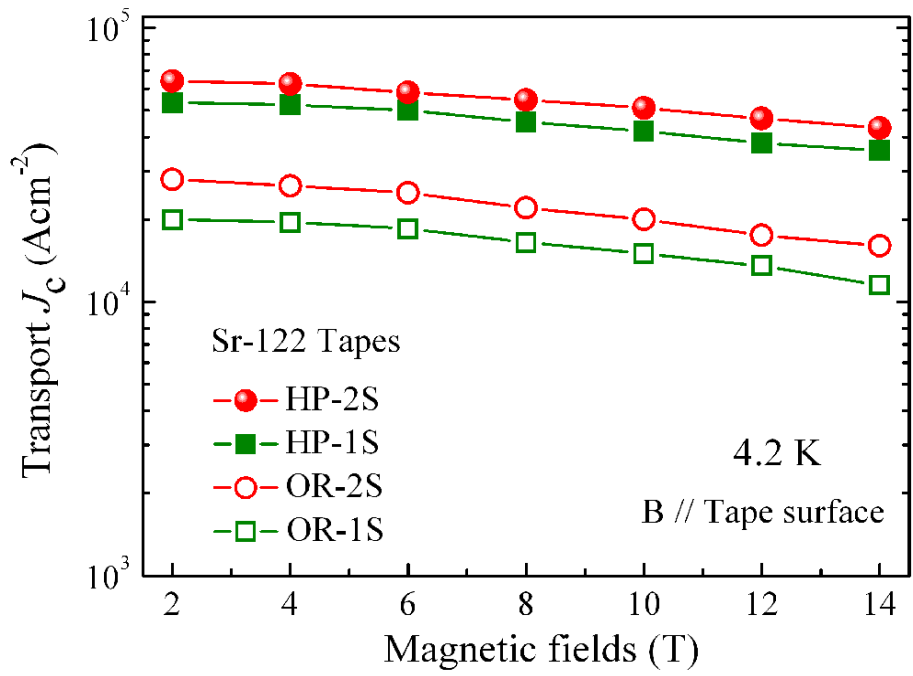

Figure 6 Lin et al.

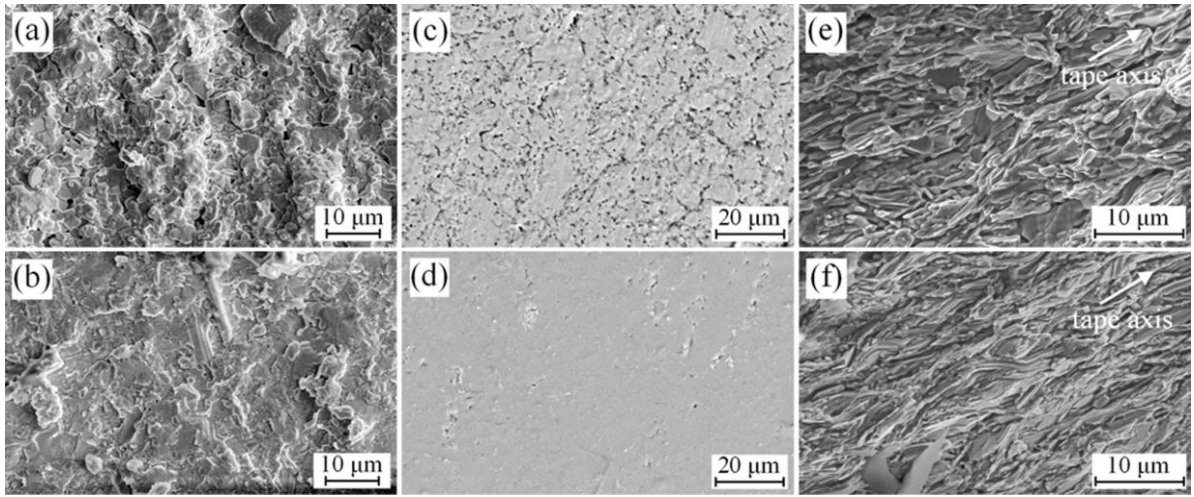

Figure 7 Lin et al.